# Reversible magnetization measurement of the anisotropy of the London penetration depth in MgB$_2$ single crystals

Heon-Jung Kim,[a)] Byeongwon Kang, Min-Seok Park, Kyung-Hee Kim, Hyun Sook Lee, and Sung-Ik Lee[b)]

*National Creative Research Initiative Center for Superconductivity and Department of Physics,*

*Pohang University of Science and Technology, Pohang 790-784, Republic of Korea*

**Abstract**

We have studied the anisotropy of the London penetration depth, which was obtained from reversible magnetization measurements with the magnetic field both parallel and perpendicular to the *c*-axis. The anisotropy of the London penetration depth has a smaller magnitude than the anisotropy of the upper critical field and increases with temperature while that of the upper critical field decreases as reported earlier. This behavior is in sharp contrast with the behaviors of superconductors with one superconducting energy gap. The temperature dependence of the anisotropies of the London penetration depth and of the upper critical field can be well explained within the theory of two-gap superconductivity in MgB$_2$.

# I. Introduction

Since the discovery of superconductivity in MgB$_2$ [1], that material has drawn great attention in the scientific community [2-7]. One reason that makes MgB$_2$ very interesting is its multi-gap property. It is now well established that MgB$_2$ has two different superconducting energy gaps [5, 6, 8-23]: a large gap originating from two-dimensional (2D) $\sigma$ bands and a small gap originating from three-dimensional (3D) $\pi$ bands. Thus, any physical properties related to the superconductivity should be influenced by this multi-gap property. Among those, a strong temperature dependence has been observed experimentally for the anisotropy of the upper critical field $H_{c2}$ ($\gamma_H \equiv \xi_{ab}/\xi_c = H_{c2//ab}/H_{c2//c}$), which decreases from around 5 at low temperatures to 2 near $T_c$ [24-27]. According to the one-gap Ginzburg-Landau theory, this is totally unexpected because the anisotropy defined as $\gamma \equiv \gamma_\lambda = \gamma_H$ should be constant. Therefore, this temperature dependence of $\gamma_H$ is thought to result from the interplay of two superconducting energy gaps, and several theoretical works have revealed that two different gaps affect the behavior of $\gamma_H$ [28-30]. For example, Dahm and Schopohl calculated $H_{c2}$ in the clean limit based on a detailed modeling of the electronic structure that took into account the Fermi surface topology and the two-gap nature of the order parameter [28] while Gurevich studied $H_{c2}$ in the dirty limit by using quasi-classical Usadel equations for two-band superconductivity [29]. The results show that $\gamma_H$ depends on the interplay of two different superconducting gaps in both the clean and the dirty limits; however, in the dirty limit, the ratio of the intraband electron diffusivities, which depends on the impurity level, becomes important.

Another consequence of the multi-gap nature of $MgB_2$ is that the anisotropy of the penetration depth ($\gamma_\lambda$), as well as the anisotropy of $H_{c2}$ may no longer be described by a single parameter [31-34]. In $MgB_2$, since the penetration depth depends on the total number of charge carriers from both the σ band and the π band while $H_{c2}$ is mainly determined by the σ band, $\gamma_\lambda$ is not necessarily the same as $\gamma_H$ [31, 32]. According to Kogan's calculation based on weak-coupling theory [31], $\gamma_\lambda$ is isotropic at low temperature and increases to 2.5 near $T_c$. That behavior was confirmed by the first-principle calculations for the electronic structure and the electron-phonon interaction [32], which showed that the effect of impurity scattering changed the exact value of $\gamma_\lambda$. However, that result was not verified experimentally until recently when an estimate of $\gamma_\lambda$ in small-angle neutron-scattering (SANS) measurements on $MgB_2$ polycrystals [33] first revealed an isotropic $\gamma_\lambda$ at 2 K. Nevertheless, the temperature dependence of $\gamma_\lambda$ could not be precisely investigated due to the random nature of the polycrystalline sample. Later, using $MgB_2$ single crystals the same SANS technique was employed to measure the behavior of $\gamma_\lambda$ as a function of the temperature as well as a function of the field [34] over limited ranges of the temperature and the field. Even though the value of $\gamma_\lambda$ determined from SANS measurements increases with temperature as predicted in the calculations, the values of $\gamma_\lambda$ are larger than the theoretical values.

Values of $\gamma_\lambda$ determined from *M-H* loop measurements have also been reported. Caplin *et al.* obtained nearly equal and constant anisotropies with $\gamma_H = \gamma_\lambda \sim 2$ [35], which differed from the theoretical predictions. More recently, using a Hall sensor, Lyard *et al.* carefully investigated the temperature dependences of $\gamma_\lambda$ and $\gamma_H$ [36]. Contrary to the results of Caplin *et al.*, Lyard *et al.* found that $\gamma_\lambda$ and $\gamma_H$ showed opposite behaviors as predicted by the theory; they claimed that this is due to the two-gap effect.

In this research, we investigated the temperature dependences of the anisotropies of the penetration depth ($\gamma_\lambda$), the lower critical field ($\gamma_{H_{c1}}$), and the upper critical field ($\gamma_H$) by measuring both the reversible and the irreversible magnetizations (*M-H* loops). The reversible magnetization was analyzed based on the London model. Compared with *M-H* loop measurements for low fields, our analysis is free from the effect of the demagnetization factor and from the surface and the geometrical barriers, so the values of $\gamma_\lambda(T)$ are believed to be more reliable. Using the reported values of $\lambda_{ab}(T)$, we could determine $\lambda_c(T)$ and, hence, $\gamma_\lambda(T)$ in the temperature range of 20 K ≤ *T* ≤ 27 K. Since this analysis was possible only for a limited temperature range, we supplemented it by carefully determining values of $\gamma_{H_{c1}}$ from *M-H* loops in a way similar to the one used by Lyard *et al.*[36] but with different criteria. From those measurements of anisotropies, both $\gamma_{H_{c1}}$ and $\gamma_\lambda$ were found to be smaller than $\gamma_H$ and to increases with temperature while $\gamma_H$ was found to decrease with temperature. Together with previous experimental results, our data provide strong experimental support for the theoretical calculations based on a weakly coupled two-band superconductor.

## II. Experiment

Single crystals were grown by using a high pressure technique, which is explained in detail in previous reports [37, 38]. Two sets of single crystals were investigated by using magnetization measurements. In the first set, 10 relatively hexagonally shaped single crystals, with typical dimensions of 200 x 100 x 25 μm³, were collected on a substrate without an appreciable magnetic background with their *c*-axes aligned perpendicular to the substrate surface. In the second set, a shiny and flat single crystal with dimensions of 800 x 300 x 60 μm³ was mounted on a substrate with its *c* axis perpendicular to the substrate surface. The values of the transition temperature $T_c$ and the transition width $\Delta T_c$ determined from the low-field magnetization were 36.8 K and 1.5 K, respectively for the first set and 37.9 K and 0.7 K for the second set. Even though the values of $T_c$ were slightly different for the two sets of crystals, no other significant differences were observed during further magnetization analysis. Therefore, we only present the data for the second set in this paper.

## III. Results and Discussion

The measurements of the reversible and the irreversible magnetizations were carried out by using a superconducting quantum interference device magnetometer (Quantum Design, MPMS-XL). Figure 1 shows the temperature dependence of the reversible magnetization, $4\pi M(T)$, measured in the field range 0.5 T ≤ *H* ≤ 1.7 T with H//ab. The reversible region was determined in the temperature ranges at which the criterion $M_{FC}/M_{ZFC} \geq 0.95$ holds [39]. The reversible curves shifted to lower temperature as the field was increased. In the low magnetic fields below 0.5 T, the reversible region is too narrow to study the magnetization. On the other hand, at high fields, the magnetization itself is so small and the noise to signal ratio becomes high. Due to these reasons, the field range 0.5 T ≤ *H* ≤ 1.7 T was selected to obtain most reliable data. In our previous report [40], the reversible magnetization for H//c was analyzed using the Hao-Clem model [41]. Since for H//c, the applied magnetic fields are comparable with $H_{c2//c}(0)$, the contribution of the core energy must be significant. Therefore, only the Hao-Clem model can give proper results. However, for H//ab, the simpler London model can be utilized because $H_{c2//ab}(0)$ is much larger than the applied magnetic fields; therefore, the contribution of the core energy may not be as important as it is for the case of H//c. Later, we will show this by comparing differently calculated values of $\lambda(T)$.

According to Kogan [42], the free energy of a uniaxial superconductor for which the anisotropy of the upper critical field, $\gamma_H = H_{c2//ab}/H_{c2//c} = \xi_{ab}/\xi_c$, is different from the anisotropy of the penetration depth, $\gamma_\lambda = \lambda_c/\lambda_{ab}$, is given by

$$F = \frac{\phi_0 B \Theta_\lambda}{32\pi^2 \lambda_{ab}^2} \ln\left(\frac{2\sqrt{3}\gamma_H^{-2/3}\phi_0 \Theta_\lambda}{\xi^2 B(\Theta_\lambda + \Theta_H)^2}\right), \quad (1)$$

where $\Theta_{\lambda,H}(\theta) = \left(\sqrt{\sin^2\theta + \gamma_{\lambda,H}^2 \cos^2\theta}\right)/\gamma_{\lambda,H}$, $\phi_0$ is the flux quantum, $\lambda_{ab}$ is the in-plane penetration depth, and $\theta$ is the angle between the c axis and the induction **B**. From this free energy, the torque was calculated, and due to $\Theta_{\lambda,H}(\theta)$, the equilibrium orientation of the crystal relative to the applied field was found to shift to lower angles as the difference between the two anisotropies increases [42]. For H//c and H//ab, the free energy becomes

$$F(\theta = 0) = \frac{\phi_0 B}{32\pi^2 \lambda_{ab}^2} \ln\left(\frac{\sqrt{3}\gamma_H^{-2/3}\phi_0}{2\xi^2 B}\right)$$

and

$$F\left(\theta = \frac{\pi}{2}\right) = \frac{\phi_0 B}{32\pi^2 \lambda_{ab}\lambda_c} \ln\left(\frac{2\sqrt{3}\gamma_H^{-2/3}\phi_0 \gamma_\lambda^{-1}}{\xi^2 B(\gamma_\lambda^{-1} + \gamma_H^{-1})^2}\right), \quad (2)$$

respectively. These results reduce to that of the original London model when the various anisotropies are equal to each other. With these equations, we calculated both the in-plane and the out-of-plane penetration depths.

From the relation $M = -\partial F / \partial H$, the magnetization can be calculated. For H//ab, the magnetization gives

$$\frac{\partial M}{\partial \ln H} = \frac{\phi_0}{32\pi^2 \lambda_{ab} \lambda_c} , \qquad (3)$$

if it is assumed that the logarithmic term in the magnetization does not change drastically. When this equation is combined with our previously reported values of $\lambda_{ab}(T)$ [40], $\lambda_c(T)$ can be determined, and the result is shown in Fig. 2. For comparison, $\lambda_{ab}(T)$ is included in the same figure. To check the model dependence of $\lambda_c(T)$, we also used the Hao-Clem model and assumed that the calculated $\lambda(T)$ for H//ab was $\lambda_{eff}(T) = \sqrt{\lambda_{ab}(T)\lambda_c(T)}$. The result of the Hao-Clem model was not so different from that of the London model, which indicates that the core-energy contribution is negligible in the field range of our experiment, as we expected. The magnitude of $\lambda_c$ is 1.5 times larger than that of $\lambda_{ab}$ at 20 K. The errors in $\lambda_c(T)$ occur during the interpolation and are calculated from the difference between the data and the theoretical curves.

A notable feature of $\lambda_{ab}(T)$ and $\lambda_c(T)$ is that the difference between $\lambda_{ab}(T)$ and $\lambda_c(T)$ increases as the temperature is increased, which implies that the anisotropy of $\lambda$ ($\gamma_\lambda \equiv \lambda_c / \lambda_{ab}$) increases as the temperature is increased. According to the theoretical predictions [31, 32], $\gamma_\lambda$ is almost isotropic at low temperatures and increases to the same value as $\gamma_H$ near $T_c$. This tendency is observed in Fig. 2.

To further investigate the anisotropy of $\lambda$ at low temperatures, we directly measured the lower critical field $H_{c1}$ by using *M-H* loops. In the *M-H* loops, $H_{c1}$ was selected from the first penetrating fields at which a deviation from Meissner shielding occurs. Since the actual field near the sample is larger than the applied field due to the demagnetization effect by a factor of $1/(1-N)$, where $N$ is the demagnetization factor, that effect should be considered in calculating $H_{c1}$. Especially, the demagnetization effect is important for H//c because the demagnetization factor is large ($N_c \approx 0.7$) in that direction. The demagnetization factor perpendicular to the $c$ axis can be obtained from the relation $N_c + 2N_{ab} = 1$ and the value $N_{ab} \approx 0.15$.

To find the first penetrating fields more exactly, we calculated the deviation from the Meissner slope, which is defined as the difference between the data and the linear interpolation ($\Delta M$) [43, 44]. For example, the $\Delta M$ curves for 5 K and 30 K are shown in the inset of Fig. 3. Actually, this criterion to determine $H_{c1}$ is different from that of Lyard *et al.* [36] who selected the first minimum point in the *M-H* loops as $H_{c1}$. However, before the first minimum point, vortices usually start to penetrate into the sample and that makes the Meissner slope deviate from a straight line. Therefore, if the minimum point is selected as $H_{c1}$, the values of $H_{c1}$ might be overestimated. Actually, the same criterion as ours is used to determine $H_{c1}$ for high-$T_c$ superconductors [43, 44]. Since Lyard *et al.* used a Hall sensor to measure the magnetization [36], this effect may not be pronounced. Nevertheless, as manifested by the factor of 2 difference between the values of $H_{c1}$ in their work and ours, their criterion may provide an upper bound on $H_{c1}$. Figure 3 displays the temperature dependence of $H_{c1}$ which was determined by using this method. Contrary to $H_{c2}$, $H_{c1}$ is almost isotropic, as reported by Lyard *et al.* [36].

To check the validity of our method for determining the values of $H_{c1}$, we estimated $\lambda_{ab}(0)$ from $H_{c1//c}(0) \approx 534$ G. The value of $\lambda_{ab}(0)$ was calculated to be 76 *nm* by using the formula $H_{c1//c} = \phi_0 / 4\pi \lambda_{ab}^2 \ln \kappa_{ab}$ with $\kappa_{ab} = 6.4$ [40]. This value of 76 *nm* is in good agreement with the values obtained from the reversible magnetization study, which guarantees the validity of our method. Encouraged with this result, we calculated both the anisotropy of $H_{c1}$ ($\gamma_{H_{c1}}$), which is defined as $H_{c1//c}/H_{c1//ab}$ and the anisotropy of $\gamma_\lambda$ that is obtained from $H_{c1}(T)$ in *M-H* loops using the reported $H_{c2}(T)$ [ref].

To compare the temperature dependences of $\gamma_\lambda$, $\gamma_{H_{c1}}$, and $\gamma_H$, we summarize all the quantities in Fig. 4. The values of $\gamma_H$ were deduced from the values of $H_{c2}$, which were determined from the onset of superconductivity in the magnetization measurements. $\gamma_\lambda$ and $\gamma_{H_{c1}}$ show a similar temperature dependence though small difference exists between the values of $\gamma_\lambda$ and $\gamma_{H_{c1}}$, which may originate from the logarithmic term in the relation between $H_{c1}$ and $\lambda$. When $\gamma_{H_{c1}}$ is converted into $\gamma_\lambda$ using $H_{c2}(T)$ in the literature [24,25], the values become very consistent with those deduced from the reversible magnetization. In striking contrast to the behavior of $\gamma_H$ [24-30] which decreases with temperature and approaches 2, $\gamma_\lambda$ and $\gamma_{H_{c1}}$ increase with temperature and show a tendency to converge to the value of $\gamma_H$ near $T_c$. Compared with the results of Lyard *et al*. [36], our values of $\gamma_{H_{c1}}$ are a little smaller. We believe that this is due to the use of different criteria for determining $H_{c1}$. Nevertheless, the overall features of the anisotropy data for $\lambda$ or $H_{c1}$ are very similar in both our measurements and theirs. According to the theoretical predictions [28-30], the decrease in $\gamma_H$ is due to different contributions from the smaller and the larger gaps at different temperatures. On the contrary, since the penetration depth depends solely on the carrier density of the $\pi$ band at low temperatures, in a clean limit, $\gamma_\lambda$ is almost isotropic and increases only weakly with temperature due to the effect of two different bands [31, 32].

Eventually, the values of $\gamma_H$ and $\gamma_\lambda$ approach the same value at $T = T_c$, and as for one-gap superconductors, a common value of the anisotropy can be determined by using the mass tensor. This overall behavior is clearly shown in Fig. 4 and our data follow the clean-limit theoretical predictions of $\gamma_\lambda$.

In fact, MgB$_2$ single crystals are believed to be a clean-limit superconductor, with $\sigma$-band probably in the clean limit [20, 46,47]. According to the theoretical calculation [32], if $\pi$ band is very impure, $\gamma_\lambda$ would increase drastically but this behavior was not observed in this study. The clean limit formalism also successfully described the temperature dependence of $\gamma_H$ [28]. Furthermore, various phenomena related to the surface superconductivity [48] as well as the peak effect [49] which appears in a very clean superconductor were observed to be pronounced in our crystals.

### IV. Summary

We have investigated the anisotropies of the penetration depth, the lower critical field, and the upper critical field. The anisotropy of the penetration depth obtained from the reversible magnetization in the temperature range 20 K $\leq T \leq$ 27 K increases as the temperature is increased. This trend is also observed in the anisotropy of the lower critical field over a wider temperature ranges, but the anisotropy of the upper critical field shows an opposite behavior. Near $T_c$, all three anisotropies approach a common value of 2. The fact that differently determined anisotropies show the same trend suggests that this property is generic in MgB$_2$. Furthermore, the temperature dependencies of the anisotropies agree well with the theoretical predictions even though the values of the anisotropies are a little lower than the predicted ones.

## ACKNOWLEDGMENT

This work was supported by the Ministry of Science and Technology of Korea through the Creative Research Initiative Program.

**Figure Captions**

Fig. 1. Temperature dependence of the reversible magnetization, $4\pi M(T)$, of a MgB$_2$ single crystal for various field ranges of H//ab in the field range of 0.5 T $\leq H \leq$ 1.7 T for the sample with $T_c$ = 37.9 K and $\Delta T_c \sim$ 0.7 K.

Fig. 2. Temperature dependence of the in-plane ($\lambda_{ab}(T)$) and out-of-plane ($\lambda_c(T)$) penetration depth. $\lambda_c(T)$ was calculated using the London model, and the $\lambda_{ab}(T)$ was taken from a previous report [38]. $\lambda_c$ is 1.5 times larger than $\lambda_{ab}$ at 20 K and increases faster.

Fig. 3. Temperature dependence of the lower critical field, $H_{c1}(T)$, for H//ab and H//c as obtained from the M-H loops. To calculate the $H_{c1}(T)$, we carefully considered the demagnetization factor of our sample. The inset shows the deviation from the Meissner slope, which is the difference between the data and linear interpolation. $H_{c1}$ was taken as the field at which $\Delta M$ starts to deviate from a constant value. For clarity, the 30 K data was shifted vertically.

Fig. 4. Temperature dependences of the anisotropies of $\lambda$ ($\gamma_\lambda$) that were obtained from reversible magnetization and direct measurements of $H_{c1}$, the anisotropy of $H_{c1}$ ($\gamma_{H_{c1}}$), and the anisotropy of $H_{c2}(\gamma_H)$. While $\gamma_H$ decreases with temperature to a value of 2 near $T_c$, both $\gamma_\lambda$ and $\gamma_{H_{c1}}$ increases. As the theory predicts, $\gamma_\lambda, \gamma_{H_{c1}}$, and $\gamma_H$ merge into the same value at $T = T_c$.

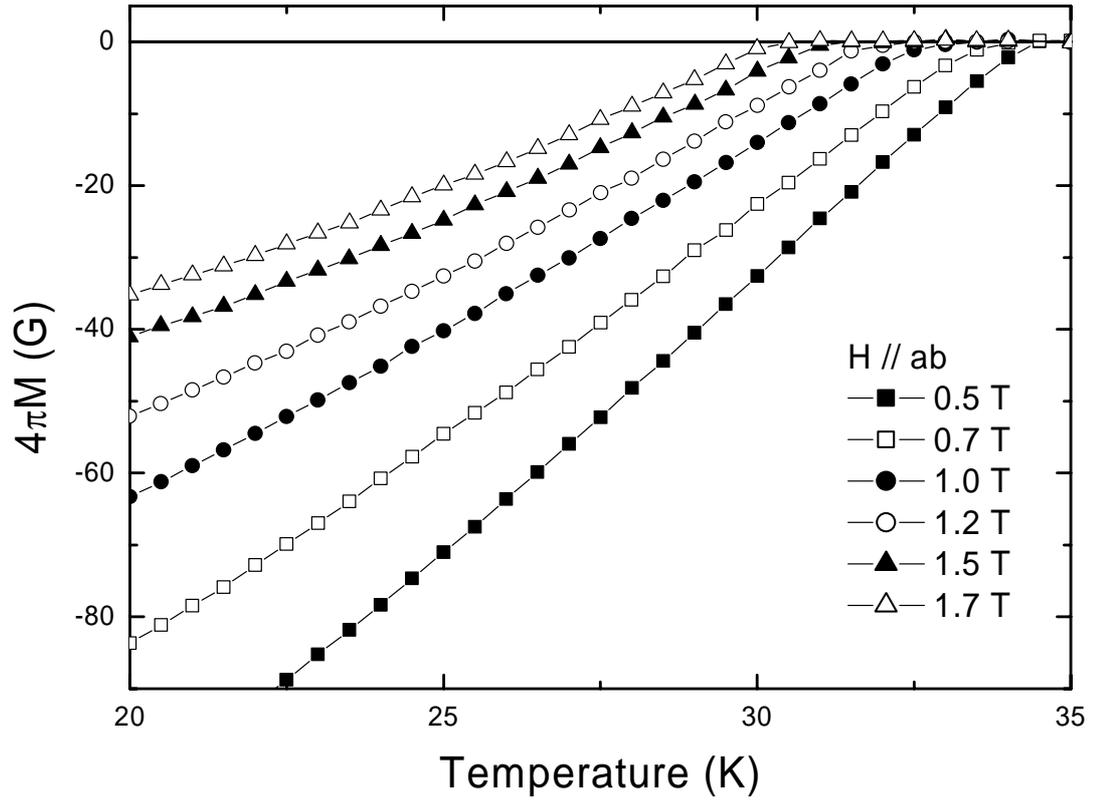

Figure 1 f Kim *et al.*

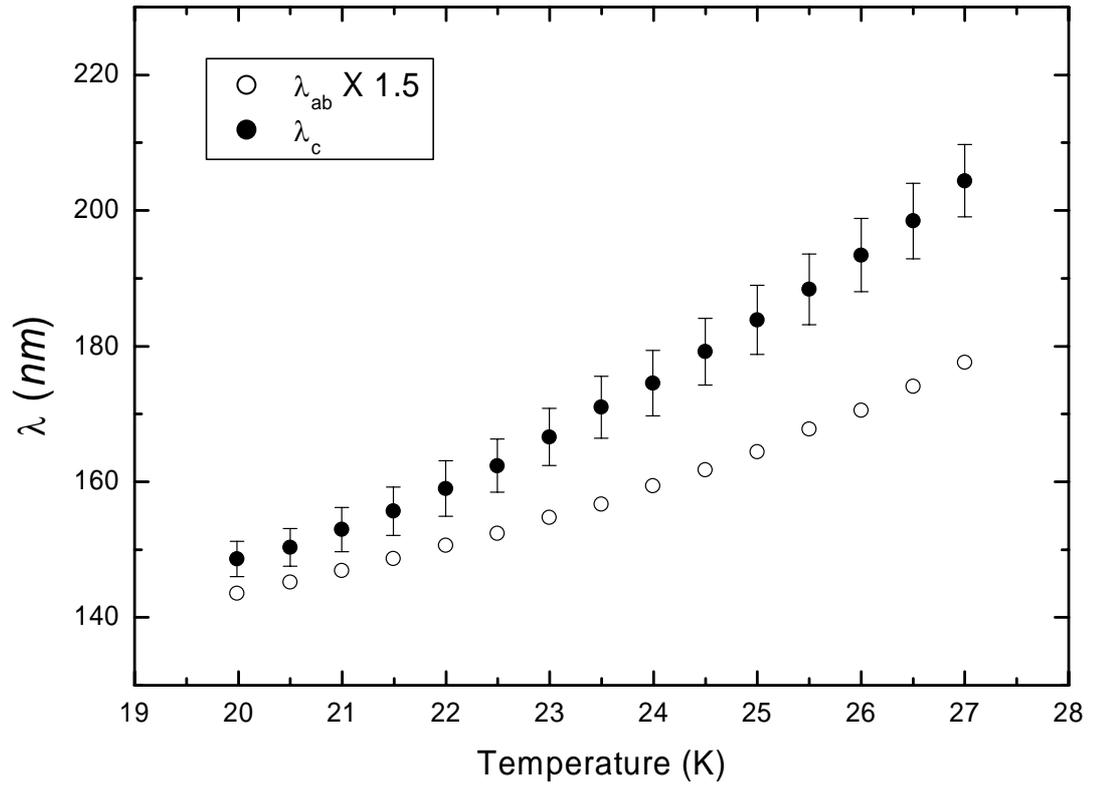

Figure 2 of Kim *et al.*

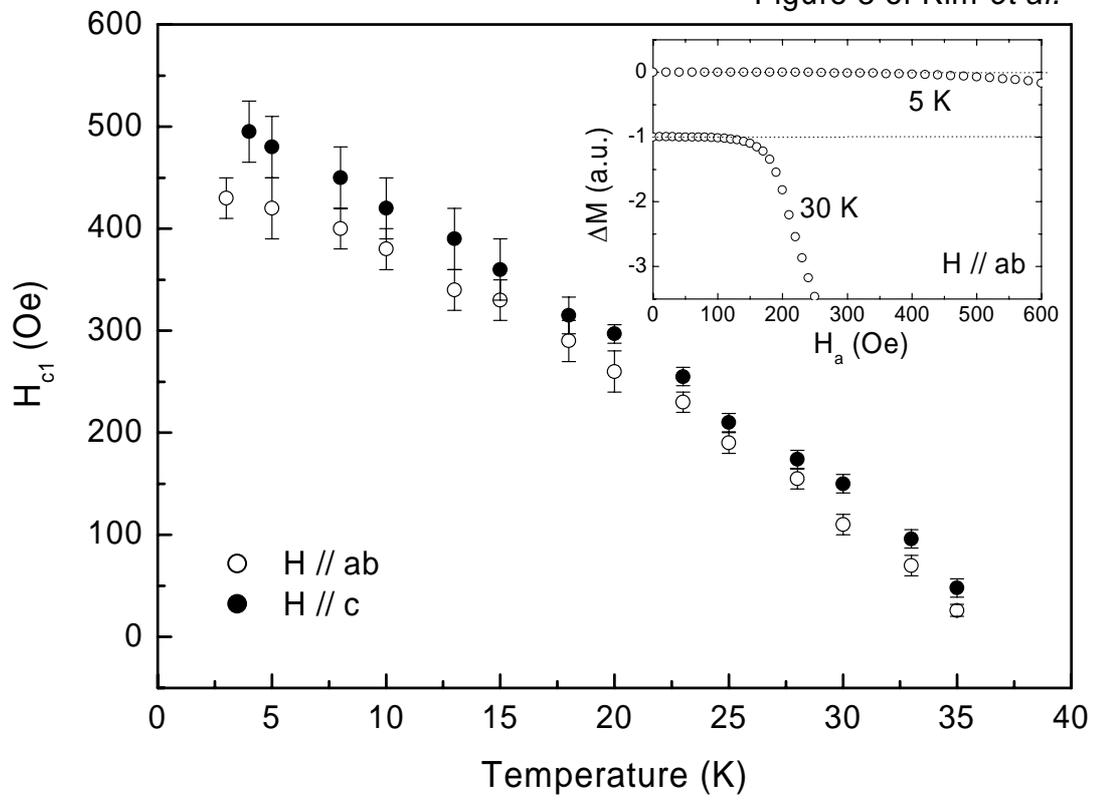

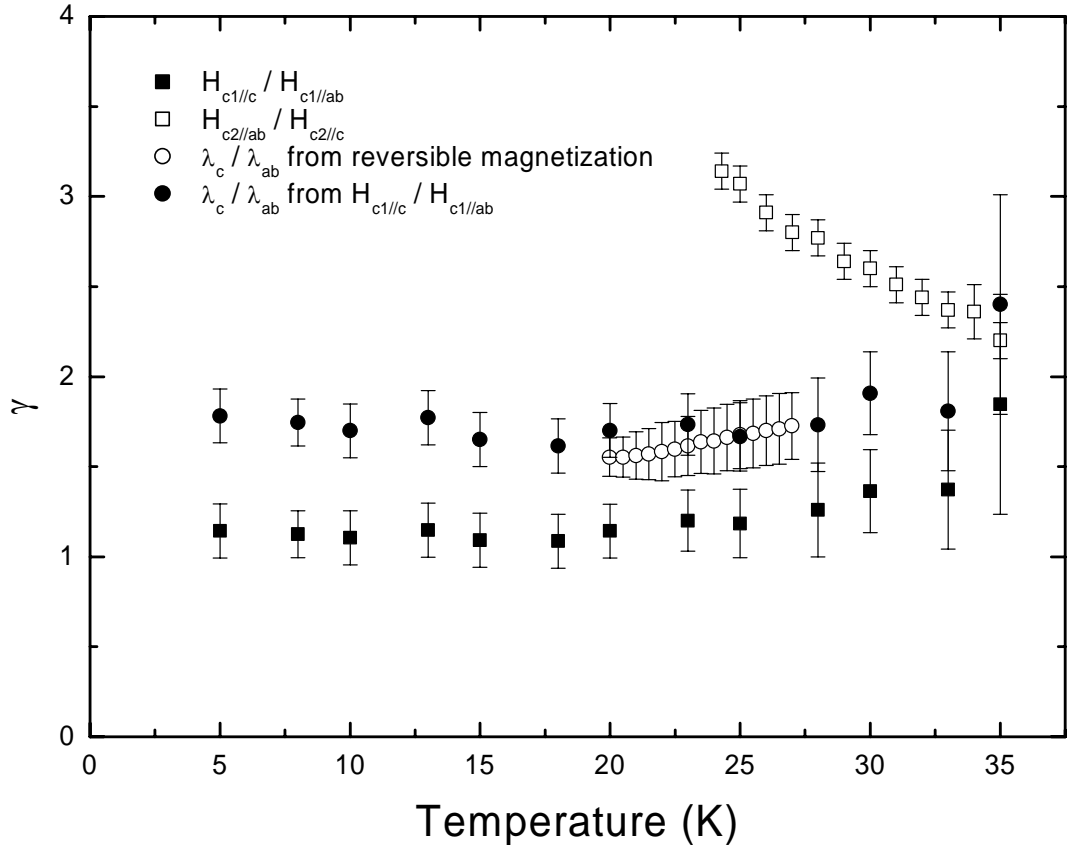

Figure 4 of Kim *et al.*